\documentclass[11pt,a4paper]{article}
\usepackage{amsmath}
\usepackage{float}
\usepackage{caption2}
\usepackage{graphicx}

\begin{document}

\title{\textbf{\large{Degree of entanglement for two qutrits in a pure state}}}
\vspace{1cm}
\author{Jos\'{e} L.\ Cereceda\thanks{Electronic mail: jl.cereceda@teleline.es} \\
\textit{C/Alto del Le\'{o}n 8, 4A, 28038 Madrid, Spain}}

\date{May 8, 2003}

\maketitle

\begin{abstract}
In this paper, a new measure of entanglement for general pure bipartite states of two qutrits is formulated.

\vspace{.2cm}
\noindent\textit{Keywords:} Qutrit, bipartite entangled state, entanglement of formation, concurrence, Schmidt decomposition.

\end{abstract}

\vspace{.5cm}

One main goal of modern quantum theory is the characterization and quantification of the property of entanglement exhibited by composite quantum systems, as entanglement is the key resource in many of the recent quantum information applications. In particular, quantum dense coding \cite{BW92}, quantum teleportation \cite{BBCJPW93}, and certain types of quantum cryptographic key distributions \cite{Ekert91,JSWWZ00}, rest crucially on the existence of entangled states. The question: ``Given a quantum state, is it separable or entangled?'', is nowadays completely solved for the simplest case of a pair of two-level systems or qubits (i.e., $2\times 2$ dimensional systems) \cite{Wootters98}. Moreover, for this case, the Peres-Horodecki criterion of positivity of the partial transposition (PPT) \cite{Peres96,Horodecki96} allows one to ascertain whether a general state of two qubits is separable or entangled.\footnote{
The Peres-Horodecki criterion also applies to composite systems of dimension $2\times 3$. For $2\times 2$ and $2\times 3$ systems, the PPT is a necessary and sufficient condition for separability \cite{Horodecki96}.}
However, concerning bipartite systems, no operational criteria have been established that allow to classify a given state as separable or entangled for higher dimensional systems. On the other hand, several entanglement measures which quantify the amount of entanglement contained in a given state have been introduced in order to answer the following question: ``Given an entangled state, how much is it entangled?'' (for a review of the main entanglement measures and other related topics see, for instance, Refs.\ \cite{VPRK97,PV98,Springer00,Bruss01}). For the special case of bipartite pure state $\rho_{AB} = |\psi\rangle\langle\psi|$, a convenient measure of the degree of entanglement is the von Neumann entropy of either of the two subsystems $A$ or $B$ \cite{BBPS96}:
\begin{equation}
E(|\psi\rangle)=-\text{Tr}(\rho_A\log_2\rho_A)=-\text{Tr}(\rho_B\log_2\rho_B),
\end{equation}
where $\rho_A$ ($\rho_B$) is the partial trace of $|\psi\rangle\langle\psi|$ over subsystem $B$ ($A$). The measure $E(|\psi\rangle)$ is also known as the entanglement of formation (EOF) of $|\psi\rangle$ \cite{BDSW96}. It can be shown  that for the case of two qubits in the general pure state
\begin{equation}
|\psi_{2\times2}\rangle = \sum_{i,j=1}^{2} \alpha_{ij} |i,j\rangle,
\end{equation}
with normalization $\sum_{ij}|\alpha_{ij}|^{2}=1$, the EOF (1) is given by \cite{Wootters98,BDSW96}
\begin{equation}
E(|\psi_{2\times2}\rangle) = h \left( \frac{1+\sqrt{1-C^2}}{2} \right),
\end{equation}
where $h$ is the binary entropy function
\begin{equation}
h(x) = -x\log_2 x-(1-x) \log_2 (1-x),
\end{equation}
and where the \textit{concurrence\/} $C$ is
\begin{equation}
C(|\psi_{2\times2}\rangle) = 2|\alpha_{11}\alpha_{22}-\alpha_{12}\alpha_{21}|.
\end{equation}
$E$ is a monotonically increasing function of $C$, and then the quantity $C$ itself can be taken as a measure of entanglement. We also note that the concurrence (5) can be written equivalently as \cite{ASST01}
\begin{equation}
C(|\psi_{2\times2}\rangle) = 2\kappa_1\kappa_2,
\end{equation}
where $\kappa_1$ and $\kappa_2$ are the two coefficients appearing in the Schmidt decomposition \cite{Peres93+EK95}, $|\psi_{2\times2}\rangle = \kappa_1 |x_1,y_1\rangle + \kappa_2 |x_2,y_2\rangle$, and $\{|x_1\rangle,|x_2\rangle\}$ and $\{|y_1\rangle,|y_2\rangle\}$ are orthonormal bases for the Hilbert spaces of subsytems $A$ and $B$, respectively.

In this paper, we formulate a measure of entanglement for a pair of three-level systems or qutrits (i.e., $3\times 3$ dimensional systems) in an arbitrary pure state. This measure is the generalization to two qutrits of the concurrence for two qubits defined in either (5) or (6). Our measure of entanglement fulfills the following properties: (i) it is unique for a given quantum state; (ii) it is invariant under local unitary operations; (iii) it ranges in the interval $[0,1]$, the value $0$ $(1)$ attained by the product (respectively, maximally entangled) state of two qutrits. Several aspects of qutrit entanglement have been previously considered by Caves and Milburn in Ref.~\cite{CM99}, where the separability of various mixed states of two or more qutrits is investigated. Furthermore, in Ref.~\cite{FCZ02} the degree of entanglement for a class of pure states of two qutrits is suggested in the context of a study of the Bell-CHSH inequality for two qutrits \cite{CGLMP02,KKCZO02,Cereceda03}. We will then compare our measure of entanglement with the one proposed in \cite{FCZ02}, and show that the present measure constitutes the proper one characterizing the degree of entanglement for a general pure state of two qutrits. We mention that our work also represents an extension to two qutrits of the degree of entanglement for two qubits formulated in Ref.~\cite{CFUZ02}.

We start by noting that any state $\rho_{AB}$ of two qutrits can be expanded uniquely as \cite{CM99}
\begin{equation}
\rho_{AB}= \frac{1}{9} \left( \textbf{1}\otimes\textbf{1} +\sqrt{3} \, \vec{\lambda}^{A}\cdot 
           \textbf{u}\otimes\textbf{1} + \sqrt{3} \, \textbf{1}\otimes\vec{\lambda}^{B}\cdot 
           \textbf{v} + \frac{3}{2} \sum_{i,j=1}^{8} \beta_{ij} \lambda_{i}^{A}
           \otimes\lambda_{j}^{B} \right),
\end{equation}
where $\lambda_i$ ($i=1,2,\ldots,8)$ are the eight Hermitian generators of SU(3). For completeness, in the Appendix we write down the matrix representations of the generators $\lambda_i$ in the standard, orthonormal basis $\{|1\rangle, |2\rangle, |3\rangle\}$ for a qutrit. Both $\textbf{u}\equiv\{u_1,\ldots,u_8\}$ and $\textbf{v}\equiv\{v_1,\ldots,v_8\}$ can be regarded as vectors in a real, eight-dimensional vector space. Likewise, $\vec{\lambda}^{A}$ and $\vec{\lambda}^{B}$ are operator vectors acting in the Hilbert spaces of subsystems $A$ and $B$, respectively. The (real) expansion coefficients in Eq.\ (7) are given by
\begin{align}
u_i = & \, \frac{\sqrt{3}}{2}\, \text{Tr}(\rho_{AB} \lambda_i \otimes\textbf{1}), \\
v_j = & \, \frac{\sqrt{3}}{2}\, \text{Tr}(\rho_{AB} \textbf{1}\otimes\lambda_j), \\
\beta_{ij} = & \, \frac{3}{2}\, \text{Tr} (\rho_{AB} \lambda_i \otimes\lambda_j).
\end{align}
It is not difficult to show that if $\rho_{AB}$ corresponds to a pure state $\rho_{AB} = |\psi\rangle\langle\psi|$ then we have that $|\textbf{u}|=|\textbf{v}|$. Furthermore, if $|\psi\rangle$ is a product state then $|\textbf{u}|=|\textbf{v}|=1$. This latter follows from the fact that the reduced density matrix of, say, subsystem $A$ is given by
\begin{equation}
\rho_A = \text{Tr}_{B}\rho_{AB}=\frac{1}{3}(\textbf{1}+\sqrt{3}\,\vec{\lambda}^{A}\cdot \textbf{u}).
\end{equation}
But, if $\rho_{AB} = |\psi\rangle\langle\psi|$ and $|\psi\rangle$ is a product state, it is obvious that $\rho_{A}^2 = \rho_A$. Applying this condition to the operator (11), we obtain that $|\textbf{u}|=1$. Note, on the other hand, that the case $|\textbf{u}|=0$ corresponds to the maximally mixed reduced density operators $\rho_A = \rho_B = \tfrac{1}{3}\textbf{1}$. This in turn implies that the original pure state $\rho_{AB} = |\psi\rangle\langle\psi|$ is maximally entangled.

Let us now consider the scalar quantity
\begin{equation}
C = \sqrt{1-|\textbf{u}|^2}. 
\end{equation}
The quantity $C$ seems to be a good candidate to measure the relative amount of entanglement contained in a pure state $\rho_{AB} = |\psi\rangle\langle\psi|$. Indeed, as we have just seen, $C=0$ for the product state whereas $C=1$ for the maximally entangled state. Admittedly, the main motivation to choose $C$ as a measure of the entanglement present in a pure state of two qutrits is that the concurrence for two qubits in Eq.\ (5) or (6) can be equally defined by $C(|\psi_{2\times2}\rangle) =\sqrt{1-|\textbf{u}|^2}$, where in this case $\textbf{u}$ is the Bloch vector determining the reduced density matrix $\rho_A = \tfrac{1}{2} (\textbf{1}+\vec{\sigma}^{A}\cdot \textbf{u})$ \cite{CFUZ02}. Accordingly, we take the expression in Eq.\ (12) as the definition of the concurrence for a pure state of two qutrits, where the vector $\textbf{u}$ is in turn given by Eq.\ (8). Thus, it can be shown that for a general pure state of two qutrits of the form
\begin{equation}
|\psi_{3\times3}\rangle = \sum_{i,j=1}^{3} \alpha_{ij} |i,j\rangle,
\end{equation}
with normalization $\sum_{ij}|\alpha_{ij}|^{2}=1$, the concurrence (12) is given by
\begin{align}
 C(|\psi_{3\times3}\rangle) = & \, \big[ 3 \big( |\alpha_{11}\alpha_{22}-\alpha_{12}\alpha_{21}|^{2} + 
|\alpha_{13}\alpha_{21}-\alpha_{11}\alpha_{23}|^{2} +
|\alpha_{12}\alpha_{23}-\alpha_{13}\alpha_{22}|^{2}  \nonumber  \\
 & \,\, + |\alpha_{12}\alpha_{31}-\alpha_{11}\alpha_{32}|^{2} +
|\alpha_{11}\alpha_{33}-\alpha_{13}\alpha_{31}|^{2} +
|\alpha_{21}\alpha_{32}-\alpha_{22}\alpha_{31}|^{2}  \nonumber  \\
 & \,\, + |\alpha_{23}\alpha_{31}-\alpha_{21}\alpha_{33}|^{2} +
|\alpha_{13}\alpha_{32}-\alpha_{12}\alpha_{33}|^{2} +
|\alpha_{22}\alpha_{33}-\alpha_{23}\alpha_{32}|^{2} \big) \big]^{\frac{1}{2}} .
\end{align}
Eq.\ (14) is the generalization to two qutrits of the concurrence for two qubits in (5). It is helpful for what follows to introduce the matrix with entries given by $\alpha_{ij}$ in (13),
\begin{equation}
\hat{\alpha} \equiv \left( \begin{array}{ccc}
\alpha_{11} & \alpha_{12} & \alpha_{13}\\
\alpha_{21} & \alpha_{22} & \alpha_{23}\\
\alpha_{31} & \alpha_{32} & \alpha_{33}
\end{array} \right).
\end{equation}
It is a nice elementary exercise to show that if $|\psi_{3\times3}\rangle$ is a product state then necessarily $\det\hat{\alpha}=0$ (please note, however, that the converse of this statement is not true). In the same way, it is readily seen that if $|\psi_{3\times3}\rangle$ is a product state then each of the nine terms appearing in Eq.\ (14) vanishes, so that $C(|\psi_{3\times3}\rangle) =0$. On the other hand, if $|\psi_{3\times3}\rangle$ is a maximally entangled state then the matrix $\hat{\alpha}$ is restricted to obey the conditions $\rho_A= \hat{\alpha}\hat{\alpha}^{\dagger} = \tfrac{1}{3}\textbf{1}$ \textit{and\/} $\rho_B= \hat{\alpha}^{\dagger}\hat{\alpha}= \tfrac{1}{3}\textbf{1}$. As a result, the coefficients $\alpha_{ij}$'s are constrained in such a way that they cause $C(|\psi_{3\times3}\rangle)$ to attain its maximum value $C(|\psi_{3\times3}\rangle)=1$. We note, incidentally, that from the condition $\hat{\alpha}\hat{\alpha}^{\dagger} = \tfrac{1}{3}\textbf{1}$ or $\hat{\alpha}^{\dagger}\hat{\alpha}= \tfrac{1}{3}\textbf{1}$ one obtains immediately the result that, for the maximally entangled state, $|\det\hat{\alpha}|^2 =\tfrac{1}{27}$. Of course, for an arbitrary state the concurrence (14) lies in the interval $0\leq C(|\psi_{3\times3}\rangle) \leq 1$.

The state in (13) can be written alternatively in terms of a Schmidt decomposition \cite{Peres93+EK95}
\begin{equation}
|\psi_{3\times3}\rangle = \kappa_1 |x_1,y_1\rangle + \kappa_2 |x_2,y_2\rangle
  + \kappa_3 |x_3,y_3\rangle,
\end{equation}
where $\{|x_i\rangle\}$ and $\{|y_i\rangle\}$ are two orthonormal basis sets belonging to the Hilbert spaces of subsystems $A$ and $B$, respectively. $\kappa_1$, $\kappa_2$, and $\kappa_3$ are real and nonnegative coefficients satisfying $\kappa_1^2+\kappa_2^2+\kappa_3^2=1$ (without loss of generality we may also take $\kappa_1\geq\kappa_2\geq\kappa_3$). Since $\kappa_1$, $\kappa_2$, and $\kappa_3$ are unique for any given state $|\psi_{3\times3}\rangle$, it should be possible to uniquely express $C(|\psi_{3\times3}\rangle)$ in terms of the Schmidt coefficients $\kappa_1$, $\kappa_2$, and $\kappa_3$ for the state $|\psi_{3\times3}\rangle$. From the theory of the Schmidt decomposition \cite{Peres93+EK95}, we know that the squares $\kappa_1^2$, $\kappa_2^2$, and $\kappa_3^2$ correspond to the three eigenvalues of the reduced matrix $\rho_A= \hat{\alpha}\hat{\alpha}^{\dagger}$ or $\rho_B= \hat{\alpha}^{\dagger}\hat{\alpha}$. These eigenvalues are solutions of the cubic equation
\begin{equation}
\lambda^3 -\lambda^2 + \frac{1}{3}C^2(|\psi_{3\times3}\rangle)\lambda -|\det\hat{\alpha}|^2 = 0,
\end{equation}
with $C(|\psi_{3\times3}\rangle)$ being the expression in Eq.\ (14). The roots of the equation in (17) are real and nonnegative. Denoting such roots by $\kappa_1^2$, $\kappa_2^2$, and $\kappa_3^2$, it follows from the coefficients of the cubic equation (17) that its roots satisfy the relations:
\begin{align}
& \kappa_1^2+\kappa_2^2+\kappa_3^2=1,  \\
& \kappa_1^2\kappa_2^2 + \kappa_1^2\kappa_3^2 + \kappa_2^2\kappa_3^2 = 
             \frac{1}{3}C^2(|\psi_{3\times3}\rangle),  \\
& \kappa_1^2 \kappa_2^2 \kappa_3^2= |\det\hat{\alpha}|^2.
\end{align}
From Eq.\ (19), we obtain the concurrence $C(|\psi_{3\times3}\rangle)$ in terms of the Schmidt coefficients associated with $|\psi_{3\times3}\rangle$
\begin{equation}
C(|\psi_{3\times3}\rangle) =\sqrt{3 (\kappa_1^2\kappa_2^2 + \kappa_1^2\kappa_3^2 
 + \kappa_2^2\kappa_3^2 )}.
\end{equation}
Eq.\ (21) is the generalization to two qutrits of the concurrence for two qubits in (6). As we have seen, for the maximally entangled state we have $C(|\psi_{3\times3}\rangle)=1$ and $|\det\hat{\alpha}|^2 =\tfrac{1}{27}$. Substituting these values into Eqs.\ (18)-(20) gives $\kappa_1^2=\kappa_2^2=\kappa_3^2 =\tfrac{1}{3}$. On the other hand, for product states we have $C(|\psi_{3\times3}\rangle)=0$ and $\det\hat{\alpha}=0$, which gives $\kappa_1^2=1$, $\kappa_2^2=\kappa_3^2=0$. The case in which $C(|\psi_{3\times3}\rangle)\neq 0$ and $\det\hat{\alpha}=0$, corresponds to the states with $\kappa_1^2, \kappa_2^2\neq 0$ and $\kappa_3^2=0$. When one of the Schmidt coefficients (say $\kappa_3$) is zero, the concurrence (21) reduces to $C(|\psi_{3\times3}(\kappa_3=0)\rangle)\linebreak =\sqrt{3}\kappa_1\kappa_2$, ranging in the interval $[0,\tfrac{1}{2} \sqrt{3}]$. Therefore, the EOF for the class of states \linebreak\mbox{$|\psi_{3\times3}(\kappa_3=0)\rangle$} is
\begin{equation}
E(|\psi_{3\times3}(\kappa_3=0)\rangle) = h\left( \frac{1+\sqrt{1-(4C^2/3)}}{2} \right) .
\end{equation}
$E(|\psi_{3\times3}(\kappa_3=0)\rangle)$ ranges from $0$ for the product state (i.e., $C=0$) to 1 ebit for the ``singlet'' state $|\psi_{3\times3}\rangle = (1/\sqrt{2})(|x_1,y_1\rangle +|x_2,y_2\rangle)$ (i.e., $C=\sqrt{3}/2$). Naturally, we know that the maximum EOF is obtained for the maximally entangled state $|\psi_{3\times3}\rangle = (1/\sqrt{3})(|x_1,y_1\rangle +|x_2,y_2\rangle) +|x_3,y_3\rangle)$, this maximum value being equal to $\log_2 3\approx 1.58$ ebits. Unfortunately, however, we lack a general formula for the entanglement of formation $E(|\psi_{3\times3}\rangle)$ of an arbitrary pure state $|\psi_{3\times3}\rangle$ in terms of the concurrence (21).

Of course, the entanglement measure (21) satisfies the property of invariance under local unitary operations. This follows from the facts: (i) $\kappa_1^2$, $\kappa_2^2$, and $\kappa_3^2$ are the eigenvalues of either the matrix $\rho_A= \hat{\alpha}\hat{\alpha}^{\dagger}$ or $\rho_B= \hat{\alpha}^{\dagger}\hat{\alpha}$; and (ii) The eigenvalues of a matrix are unaltered if the matrix is transformed by a similarity transformation. In particular, this latter applies to local unitary transformations $U_A\rho_A U_A^{\dagger}$ and $U_B\rho_B U_B^{\dagger}$, where $U_A$ and $U_B$ stand for general $3\times 3$ unitary matrices acting in the Hilbert spaces of subsystems $A$ and $B$, respectively.

We conclude by recalling the degree of entanglement for two qutrits proposed by Fu \textit{et al.} \cite{FCZ02}, and then comparing their measure with ours. Fu \textit{et al.} considered the class of states of two qutrits in the (Schmidt) form
\begin{equation}
|\psi\rangle = \frac{1}{\sqrt{3}}\, (a_1 |1,1\rangle + a_2 |2,2\rangle
  + a_3 |3,3\rangle ),
\end{equation}
where the $a_i$'s are real coefficients varying in $[-\sqrt3,\sqrt{3}]$, and satisfying $a_1^2+a_2^2+a_3^2 =3$. By studying the Bell-CHSH inequality for two qutrits (with two von Neumann measurements per site) \cite{CGLMP02,KKCZO02,Cereceda03}, $S\leq 2$, where the specific form of the quantity $S$ is given in Eq.\ (5) of \cite{FCZ02}, \mbox{Fu \textit{et al.}} showed that, for the class of states (23), the quantum-mechanical prediction of $S$ acquires the form
\begin{equation}
S_{\text{QM}}(|\psi\rangle) = a_1a_2 T_{12}+a_1a_3 T_{13}+a_2a_3 T_{23},
\end{equation}
where the $T_{ij}$'s depends solely on the local measurement settings. Furthermore, Fu \textit{et al.} proved that the maximum value of $S_{\text{QM}}(|\psi\rangle)$ is determined by the absolute value of the cross products $a_1a_2$, $a_1a_3$, and $a_2a_3$. In view of this, Fu \textit{et al.} suggested to describe the degree of entanglement for the two-qutrit state (23) by
\begin{figure}[t]
\vspace{-1.1cm}
\centering
\captionstyle{centerlast}
\hspace{-.5cm}
\includegraphics[width=4in]{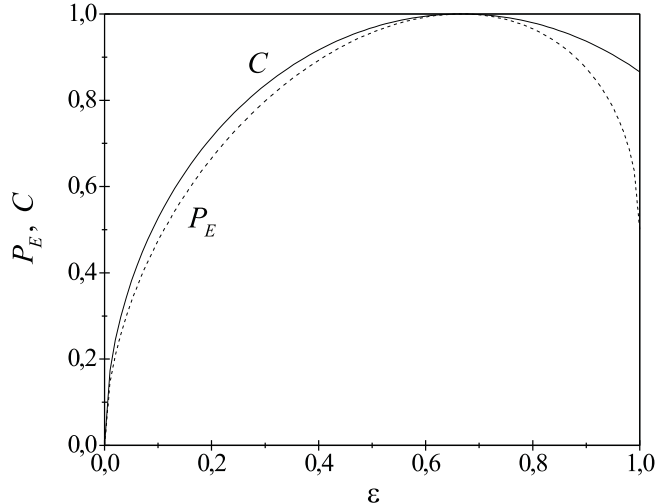}
\setlength{\abovecaptionskip}{-.4cm}
\setlength{\textfloatsep}{9pt plus 2pt minus 3pt}
\setcaptionmargin{1cm}
\renewcommand{\figurename}{Fig.}
\renewcommand{\captionlabeldelim}{.~}
\caption{\small{Degree of entanglement of the state (23) for the case in which $a_1=a_2$, where $P_E$ and $C$ are defined by Eqs.\ (25) and (26), respectively. For $a_1=a_2$, the state (23) is completely characterized by a single parameter $\epsilon$.}}
\centering
\end{figure}
\begin{equation}
P_E (|\psi\rangle) = \frac{|a_1a_2|+|a_1a_3|+|a_2a_3|}{3}.
\end{equation}
On the other hand, for the state (23), the concurrence (14) reads
\begin{equation}
C(|\psi\rangle) = \left( \frac{|a_1a_2|^2+|a_1a_3|^2+|a_2a_3|^2}{3} \right)^{\frac{1}{2}}.
\end{equation}
In Fig.\ 1, the functions $P_E (|\psi\rangle)$ and $C(|\psi\rangle)$ have been plotted for the case in which $a_1=a_2=\sqrt{3\epsilon/2}$ and $a_3=\sqrt{3(1-\epsilon)}$, where $0\leq\epsilon\leq 1$. We can see that $C\geq P_E$ for the entire range of variation of $\epsilon$. For $\epsilon=0$ (i.e., for the product state) we have $P_E=C=0$, whereas for $\epsilon=\tfrac{2}{3}$ (i.e., for the maximally entangled state) we have $P_E=C=1$. The case $\epsilon=1$ corresponds to the singlet state $|\psi_s\rangle = (1/\sqrt{2})(|1,1\rangle +|2,2\rangle)$, and for this case $P_E=\tfrac{1}{2}$ and $C=\tfrac{1}{2}\sqrt{3}$. It is to be noted that the value $P_E=\tfrac{1}{2}$ is too low to adequately describe the relative amount of entanglement present in the state $|\psi_s\rangle$, as such value leads to an underestimated entanglement of formation of the singlet state. Indeed, substituting $C$ by $\tfrac{1}{2}$ in Eq.\ (22) would give $E(|\psi_s\rangle)\approx 0.44$ ebits. However, we know that the EOF of $|\psi_s\rangle$ is, by definition, equal to $1$ ebit \cite{BDSW96}. We thus deduce that the correct measure of entanglement over the entire range $0\leq\epsilon\leq 1$ (and, in fact, for any state (23)) is the function in Eq.~(26).

To finish, we consider the generalization of the concurrence in Eq.\ (21) to the case of a pair of $d$-dimensional systems or \textit{qudits\/} ($d\geq 2$) in a pure state. Let us assume that the state of the two qudits is written in the Schmidt form
\begin{equation}
|\psi_{d\times d}\rangle = \kappa_1 |x_1,y_1\rangle + \kappa_2 |x_2,y_2\rangle
  + \cdots + \kappa_d |x_d,y_d\rangle,
\end{equation}
where $\{|x_i\rangle\}$ and $\{|y_i\rangle\}$ are orthonormal bases for the Hilbert spaces of subsystems $A$ and $B$, and the $\kappa_i$'s are real and nonnegative coefficients satisfying $\kappa_1^2+\kappa_2^2+ \cdots +\kappa_d^2=1$. Then the concurrence for the state (27) could be defined by
\begin{equation}
C(|\psi_{d\times d}\rangle) = \sqrt{\left( \frac{2d}{d-1} \right)
(\kappa_1^2\kappa_2^2+\cdots+\kappa_1^2\kappa_d^2
  + \kappa_2^2\kappa_3^2+\cdots+\kappa_2^2\kappa_d^2 + \cdots + \kappa_{d-1}^2\kappa_{d}^2) }.
\end{equation}
Note that expression (28) reduces to the concurrences (6) and (21) for the cases $d=2$ and $d=3$, respectively. Moreover, $C(|\psi_{d\times d}\rangle) =0$ for the product state, whereas $C(|\psi_{d\times d}\rangle) =1$ for the maximally entangled state, i.e. the state in (27) for which $\kappa_i=1/\sqrt{d}$ for each $i=1,\ldots,d$.

In summary, in this paper we have presented a new measure to quantify the degree of entanglement for two qutrits in a general pure state $|\psi_{3\times3}\rangle$. This measure has been expressed in two equivalent forms: either in terms of the expansion coefficients $\alpha_{ij}$ of $|\psi_{3\times3}\rangle$ in an arbitrary basis $|i,j\rangle$, $i,j=1,2,3$ (cf.~Eq.\ (14)); or else in terms of the coefficients $\kappa_i$ of $|\psi_{3\times3}\rangle$ in the Schmidt basis $\{|x_i,y_i\rangle\}$, $i=1,2,3$ (cf.~Eq.\ (21)). We have derived the entanglement of formation of $|\psi_{3\times3}\rangle$ for the simple case in which one of the Schmidt coefficients is zero. Also we have compared our measure of entanglement for two qutrits with the one previously proposed in Ref.~\cite{FCZ02}. Finally, we have extended the concurrence for two qutrits (21) to the case of two qudits described by the pure state (27).

\section*{Appendix}

In the orthonormal basis $\{|1\rangle, |2\rangle, |3\rangle\}$, the generators of SU(3) have the matrix representations:
\begin{align}
\lambda_1 & = \left( \begin{array}{ccc}
0 & 1 & 0\\ 1 & 0 & 0\\ 0 & 0 & 0 \end{array} \right),  \qquad
\lambda_2  = \left( \begin{array}{ccc}
0 & -i & 0\\ i & 0 & 0\\ 0 & 0 & 0 \end{array} \right),  \qquad
\lambda_3  = \left( \begin{array}{ccc}
1 & 0 & 0\\ 0 & -1 & 0\\ 0 & 0 & 0 \end{array} \right),  \nonumber  \\
\lambda_4 & = \left( \begin{array}{ccc}
0 & 0 & 1\\ 0 & 0 & 0\\ 1 & 0 & 0 \end{array} \right),   \qquad
\lambda_5  = \left( \begin{array}{ccc}
0 & 0 & -i\\ 0 & 0 & 0\\ i & 0 & 0 \end{array} \right),       
     \tag{A1}  \\
\lambda_6 & = \left( \begin{array}{ccc}
0 & 0 & 0\\ 0 & 0 & 1\\ 0 & 1 & 0 \end{array} \right),   \qquad
\lambda_7  = \left( \begin{array}{ccc}
0 & 0 & 0\\ 0 & 0 & -i\\ 0 & i & 0 \end{array} \right),   \qquad
\lambda_8  = \frac{1}{\sqrt{3}}\left( \begin{array}{ccc}
1 & 0 & 0\\ 0 & 1 & 0\\ 0 & 0 & -2 \end{array} \right).  \nonumber
\end{align}
The generators $\lambda_i$ obey characteristic commutation and anticommutation relations (see, for example, Refs. \cite{CM99,AMM97,ALP00}). Here we merely note that the matrices (A1) are traceless and satisfy
\begin{equation}
\text{Tr}(\lambda_\alpha\lambda_\beta) = 2 \delta_{\alpha\beta},
\qquad\alpha,\beta=0,\ldots,8,                     \tag{A2}
\end{equation}
where we have supplemented the eight generators $\lambda_i$ with the operator $\lambda_0 = \sqrt{\tfrac{2}{3}}\,\textbf{1}$ \cite{CM99}. Relation (A2) is useful in deriving the coefficients (8)-(10).

\end{document}